\documentclass[aps,prl,tightenlines,twocolumn,showpacs,preprintnumbers,amsmath,amssymb]{revtex4}
\usepackage{graphicx}
\usepackage{dcolumn}
\usepackage{bm}
\usepackage{epsfig}

\begin{document}

\title{Collapse of thermal activation in moderately damped Josephson junctions}

\author{ V.M.Krasnov$^{1,2}$, T.Bauch$^2$, S.Intiso$^2$, E.H\"{u}rfeld$^2$, T.Akazaki$^3$, H.Takayanagi$^3$, and P.Delsing$^2$}

\address{$^1$ Department of Physics, Stockholm University, Albanova University Center, SE-10691 Stockholm, Sweden\\
$^2$ Department of Microtechnology and Nanoscience, Chalmers
University of Technology, SE-41296 G\"oteborg, Sweden\\
$^3$ NTT Basic Research Laboratories, 3-1 Morinosato-Wakamiya,
Atsugi-Shi, Kanagawa 243-01, Japan}


\begin{abstract}
We study switching current statistics in different moderately
damped Josephson junctions: a paradoxical collapse of the thermal
activation with increasing temperature is reported and explained
by interplay of two conflicting consequences of thermal
fluctuations, which can both assist in premature escape and help
in retrapping back into the stationary state. We analyze the
influence of dissipation on the thermal escape by tuning the
damping parameter with a gate voltage, magnetic field, temperature
and an in-situ capacitor.

\end{abstract}
\maketitle

Decay of metastable states is an important process in many
scientific areas \cite{Hanngi}. Dissipation plays a crucial role
in the decay dynamics. The influence of dissipation on thermal and
quantum escape from the superconducting (S) to the resistive (R)
state in Josephson junctions has been intensively studied both
theoretically \cite{Hanngi,BenJacob,Grabert,Kautz} and
experimentally
\cite{Washburn,Martinis,Turlot,Vion,Silvistrini,Castellano,Han},
most recently in connection with the problem of decoherence in
quantum systems \cite{Han}.

So far switching statistics was studied for superconductor-
insulator- superconductor (SIS) junctions, while superconductor-
normal metal- superconductor (SNS) junctions, which are
characterized by stronger dissipation effects, remain, to our
knowledge, unstudied. Analysis of dissipation effects in SIS
junctions is complicated by ill-defined quality factor $Q$, which
can not be represented by a simple constant \cite{Kautz}.
Conflicting reports exists about what kind of resistance $R$
determines the effective damping of SIS junctions: the normal
resistance \cite{Washburn}, the high frequency impedance of
circuitry \cite{Martinis}, or the quasiparticle resistance
\cite{Silvistrini}. This is not a problem for SNS junctions, which
typically have a resistance $R$ considerably smaller than the open
space impedance $\simeq 377 \Omega$, and are well described by the
RCSJ model with frequency independent $Q$.

Here we study switching current statistics in moderately damped
superconductor-two dimensional electron gas-superconductor
(S-2DEG-S) and Bi$_2$Sr$_2$CaCu$_2$O$_{8+\delta}$ (Bi-2212)
high-$T_c$ intrinsic Josephson junctions (IJJ's). Being able to
tune the damping parameter by a gate voltage, magnetic field,
temperature and an in-situ shunting capacitor, we analyze the
influence of dissipation on the thermal activation (TA). For both
of the {\it drastically different systems} we observe a sudden
collapse of TA with increasing $T$ and explain this paradoxical
phenomenon by interplay of two conflicting consequences of thermal
fluctuations, which on one hand assist in premature switching and
on the other hand help in retrapping back to the S-state. We
present numeric and analytic calculations which are in good
agreement with our experimental data.

Fig. 1 shows typical Current-Voltage characteristics (IVC's) for
a) a S-2DEG-S (Nb-InAs-Nb) junction $\#1$ and c) a Bi-2212 mesa
structure containing nine IJJ's. It is seen that the IVC's are
well described by the RCSJ model with frequency independent $Q$
\cite{noteBi2212}.
Details of sample fabrication and characterization can be found in
Refs. \cite{Takayanagi,Thilo} and \cite{Bi-2212,Fluctuation} for
S-2DEG-S and IJJ's, respectively. Measurements were done in a
shielded room in a dilution refrigerator for S-2DEG-S, or in a
He-4 cryostat for Bi-2212. Switching currents were measured using
a sample-and-hold technique.

\begin{figure}
\noindent
\begin{minipage}{0.48\textwidth}
\epsfxsize=.9 \hsize \centerline{ \epsfbox{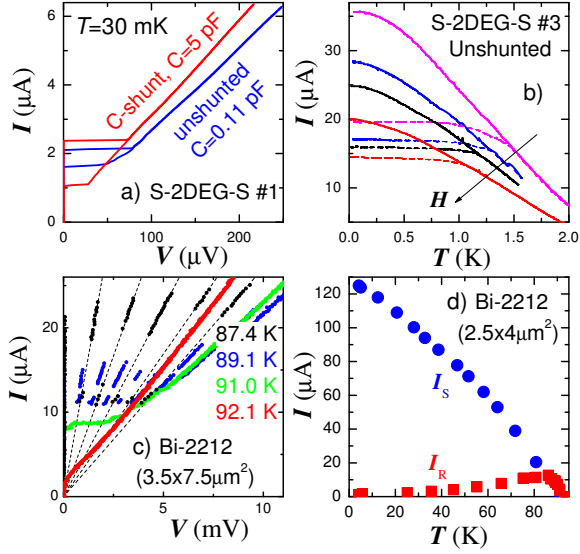} }
\caption{a) IVC's of a Nb-InAs-Nb junction $\#1$ at $T=30 mK$
before and after deposition of an in-situ shunting capacitor; b)
Measured switching (solid lines) and retrapping (dashed lines)
currents of another Nb-InAs-Nb $\#3$ at four magnetic fields. c)
IVC's of a Bi-2212 mesa containing nine stacked IJJ's at four
different $T < T_c \simeq 93K$. Multibranch structure is due to
one-by-one switching of IJJ's. d) Measured switching $I_S$ and
retrapping $I_R$ currents of another mesa. }
\end{minipage}
\end{figure}

From Fig. 1 it is seen that the IVC's exhibit hysteresis. Figs. 1
b) and d) show $T-$ dependencies of the switching, $I_S$, and
retrapping, $I_R$, currents for other Nb-InAs-Nb and Bi-2212
junctions studied here. According to the RCSJ model, $Q$ can be
obtained from the magnitude of hysteresis, $I_{S}/I_{R}$. In case
of Bi-2212, the experimental $I_S/I_R$ agrees well with the
calculated $Q$ using capacitance of IJJ's $C \simeq 68.5 fF/\mu
m^2$ \cite{Fluctuation}. For the unshunted S-2DEG-S $\#1$ from
Fig. 1 a) the $I_{S}/I_{R}\simeq 1.3$, would correspond to $Q
\simeq 1.4$ and $C=0.11 pF$, consistent with the estimated value
of the stray capacitance.

On the other hand, the hysteresis in SNS junctions can be also
caused by self-heating \cite{Fulton}, non-equilibrium effects
\cite{Song}, or frequency dependent $Q$ \cite{Kautz}. In order to
understand the origin of hysteresis, we fabricated an in-situ
shunt capacitor, consisting of Al$_2$O$_3$/Al double layer
deposited right on top of the Nb-InAs-Nb junction. The IVC's of
the S-2DEG-S $\#1$ before and after $C-$shunting are shown in Fig.
1 a). It is seen that the hysteresis increased considerably, while
$R$ was little affected by $C-$shunting. Such behavior is
inconsistent with the self-heating scenario. Thus the hysteresis
in our junctions is predominantly caused by the finite $Q>1$. A
similar conclusion was made for other planar SNS junctions
\cite{SFS}, where it was observed that there is no correlation
between the hysteresis and the dissipation power at $I=I_R$.

\begin{figure}
\noindent
\begin{minipage}{0.48\textwidth}
\epsfxsize=0.9\hsize \centerline{ \epsfbox{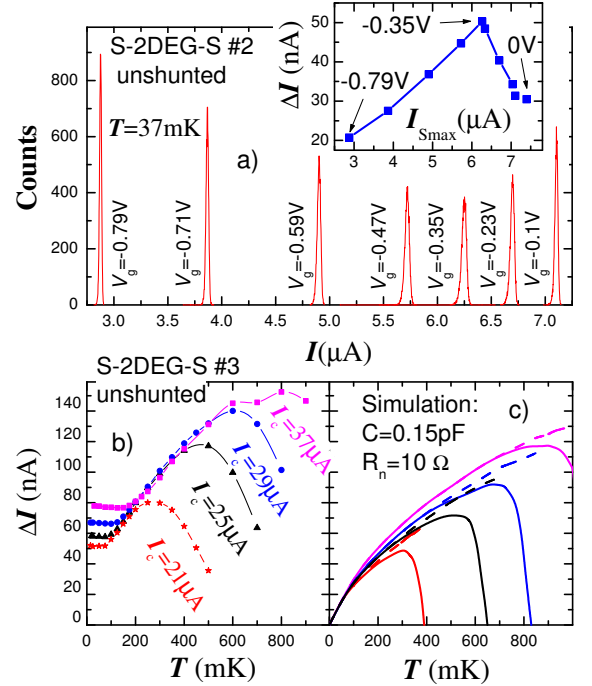} }
\caption{Switching statistics of S-2DEG-S junctions: a) histograms
at different gate voltages for junction $\#2$. Inset shows the
width vs. the most probable switching current $I_{Smax}$. A sudden
collapse of $\Delta I$ occurs at $V_g <-0.35 V$. b) The width of
histograms vs. $T$ for the junction $\#3$ at four magnetic fields,
the same as in Fig. 1b). Three $T-$regions can be distinguished:
the MQT region at low $T$, the TA region at intermediate $T$, and
collapse of histograms at higher $T$. c) Numerical simulations for
the case of Fig.2b). Dashed and solid lines represent $\Delta I$
for classical thermal activation disregarding and taking into
account retrapping, respectively.
}
\end{minipage}
\end{figure}

S-2DEG-S junctions provide a unique opportunity to tune the
Josephson coupling energy $E_{J0}$ and $Q$ by applying gate
voltage $V_g$ \cite{Takayanagi,Thilo}. For this purpose a thin
gate electrode was deposited on top of the InAs. Fig. 3 a) shows
switching current histograms at $T=37 mK$ for S-2DEG-S $\#2$ at
different $V_g$. The inset shows the width at the half-height of
the histograms $\Delta I$ vs. the most probable switching current
$I_{Smax}$. It is seen that initially histograms are getting wider
with increasing negative $V_g$, consistent with the increase of TA
with decreasing $E_{J0}/T$. However, at $V_g <-0.35 V$ a sudden
change occurs and $\Delta I$ starts to rapidly collapse.

Fig. 2 b) shows $\Delta I$ vs. $T$ for the S-2DEG-S junction $\#3$
at four magnetic fields, the same as in Fig. 1 b). In all cases we
can distinguish three $T-$ regions:

(i) At low $T$ the histograms are independent of $T$, indicative
for the Macroscopic Quantum Tunnelling (MQT) regime
\cite{Grabert,Kautz,Washburn,Martinis}. The decrease of $\Delta I$
with $H$
leaves no doubts that we observe for the first time the MQT in SNS
junctions (details will be published elsewhere).

(ii) At intermediate $T$, $\Delta I$ increases in agreement with
the TA calculations, shown by dashed lines in Fig. 2 c), for which
the escape rate from S to R state is given by
\begin{equation}
\Gamma_{TA} = a_t\frac{\omega_p}{2\pi} \exp\left[-\frac{\Delta
U}{k_B T}\right]. \label{RateTA}
\end{equation}
Here $\Delta U$ is the barrier height, and
$\omega_p=\omega_{p0}(1-(I/I_{c0})^2)^{1/4}$ is the Josephson
plasma frequency, $\omega_{p0}=(2e I_{c0}/\hbar C)^{1/2}$ and
$I_{c0}$ is the fluctuation-free critical current. Dissipation
enters only into the prefactor of Eq.(\ref{RateTA}), which for our
moderately damped junctions is $a_t=(1+1/4Q^2)^{1/2}-1/2Q$
\cite{Grabert}, where $Q=\omega_p RC$.

(iii) At higher $T$, the histograms start to rapidly collapse
leading to a downturn of $\Delta I$. This paradoxical phenomenon
is the cental observation of this work.

\begin{figure}
\noindent
\begin{minipage}{0.48\textwidth}
\epsfxsize=.9\hsize \centerline{ \epsfbox{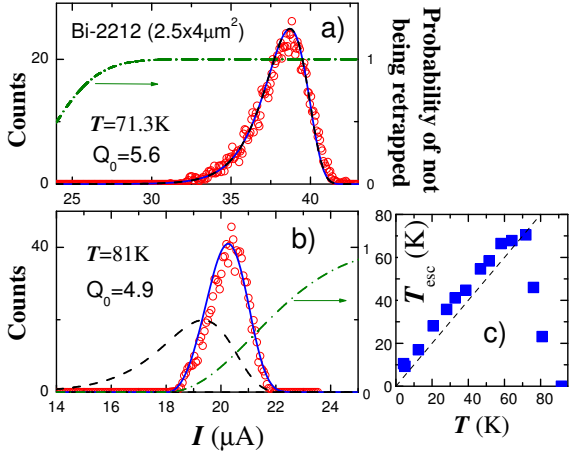} }
\caption{Switching histograms of a {\it single} IJJ at a) $T<T^*$
and b) $T>T^*$: circles represent experimental data, dashed lines
- TA simulations, Eq.(\ref{RateTA}), and dashed-dotted lines -
probabilities of not being retrapped, Eq(\ref{PnR}). Solid lines
show the conditional probability of switching without being
retrapped (the product of dashed and dashed-dotted lines). Note
that both the width and the shape of the histograms change at
$T>T^*$ c) The effective escape temperature for the same IJJ. A
sudden collapse of $T_{esc}$ at $T^*\simeq 75 K$ is seen.}
\end{minipage}
\end{figure}

Fig. 3 c) shows that a similar collapse occurs in IJJ's at $T^*
\sim 75 K$. Here we show the effective escape temperature
$T_{esc}$, which indicates how much the {\it relative} width
$\Delta I/I_{c0}$ differs from the TA prediction,
Eq.(\ref{RateTA}). Figs. 3 a) and b) show switching current
histograms of a {\it single} IJJ just before and after the
collapse. As reported previously \cite{Fluctuation}, at $T < T^*$
the histograms are perfectly described by TA, shown by the dashed
lines. However, at $T>T^*$ the histograms become narrower and
loose the characteristic asymmetric shape, as seen from comparison
with the TA simulation in Fig. 3b).

We start discussing the observed phenomenon by excluding scenarios
which can not explain it. First, it can not be due to
$T-$dependence of the damping parameter since $Q(T)$ changes only
gradually through $T^*$ and since we did take into account the
$Q-$ dependence of the TA prefactor $a_t$ in our simulations.
Second, it can not be caused by frequency dependent damping due to
shunting by circuitry impedance. Indeed, we also observed a
similar collapse for planar SNS junctions \cite{SFS} with $R\simeq
0.2 \Omega$, for which such shunting plays no role.

To explain the phenomenon we first note that $T^*$ is close to the
temperature at which hysteresis in IVC's vanishes (cf. Figs. 3c,1d
and 2b,1b), indicating that retrapping plays a role in the
observed phenomenon. The rate of TA retrapping from R to S state
is known only for strongly underdamped junctions $Q \gg 1$
\cite{BenJacob}:
\begin{equation}
\Gamma_R=\frac{I-I_{R0}}{Q_0}\sqrt{\frac{E_{J0}}{2\pi k_B
T}}\exp\left[-\frac{E_{J0}Q_0^2(I-I_{R0})^2}{2k_BT}\right],
\label{RateR}
\end{equation}
where $Q_0=\omega_{p0}RC$. Note that unlike the TA escape, the TA
retrapping depends strongly on dissipation\cite{Castellano}, due
to the $Q_0^2$ factor under the exponent in Eq.(\ref{RateR}).

The probability to measure the switching current $I$ is a
conditional probability of switching $P_S(I)$ from the S to the
R-state {\it and not being retrapped back} during the time of the
experiment:
\begin{equation}
P_{nR}=1-{\int^{I_{c0}}_I P_R(I)dI}/{\int^{I_{c0}}_0
P_R(I)dI},\label{PnR}
\end{equation}
where $P_R(I)=\frac{\Gamma_R(I)}{dI/dt} \left[1-\int^{I_{c0}}_I
P_R(I)dI\right]$ is the retrapping probability.

\begin{figure}
\noindent
\begin{minipage}{0.48\textwidth}
\epsfxsize=.9\hsize \centerline{ \epsfbox{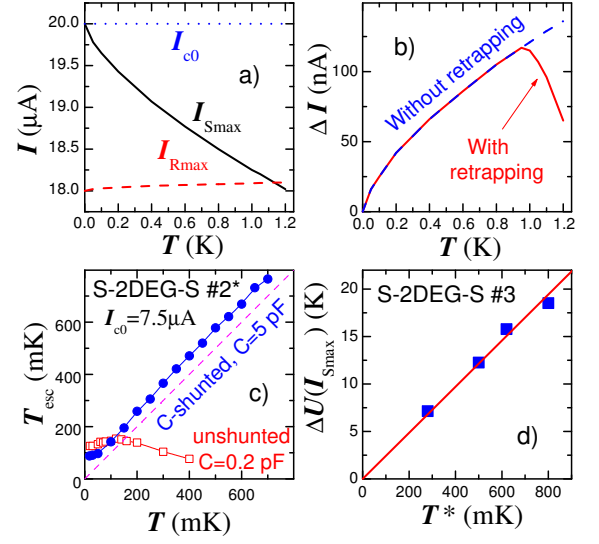} }
\caption{Numerical simulations for S-2DEG-S $\#3$: a)
$T$-dependence of the fluctuation free $I_{c0}=20\mu A$ (dotted
line), the most probable switching current $I_{Smax}$ (solid line)
and the most probable retrapping current $I_{Rmax}$ (dashed line).
b) The width of switching histograms disregarding retrapping
(dashed line) and taking into account retrapping (solid line). c)
$T_{esc}$ vs. $T$ for S-2DEG-S $\#2^*$ before and after in-situ
C-shunting. d) The height of escape barrier (at the most probable
switching current $I_{Smax}$) vs $T^*$: symbols represent
experimental data from Fig. 2b), the solid line corresponds to the
simulation in Fig. 4 b). }
\end{minipage}
\end{figure}

Dashed-dotted lines in Fig. 3 a,b) show the calculated $P_{nR}$.
It is seen that at $T<T^*$ the $P_{nR}=1$ in the region where
$P_S>0$, therefore retrapping is insignificant. However, at
$T>T^*$, retrapping becomes significant at small currents. The
resulting conditional probability of measuring the switching
current, $P(I)=P_S(I) P_{nR}(I)$, normalized by the total number
of switching events, is shown by the solid line in Fig.3 b). It is
seen that it explains very well both the reduced width and the
almost symmetric shape of the measured histogram.

Figs. 4 a,b) show results of simulations, in which we
intentionally disregarded the $T-$ dependence of $I_{c0}$, keeping
$Q_0$ and $E_{J0}$ constant. In the simulations we used the
parameters of S-2DEG-S $\#3$. It is seen that the most probable
retrapping current $I_{Rmax}$ has a weak $T-$ dependence,
consistent with the experiment, see Fig. 1 b). On the other hand,
$I_{Smax}$ decreases approximately linearly with $T$ and
eventually crosses $I_{Rmax}$. Fig. 4 b) represents the width of
histograms. It is seen that within the TA model $\Delta I$
continuously increases with $T$. However, retrapping reduces
$\Delta I$ as soon the switching and retrapping histograms start
to overlap.

The $T^*$ can be estimated from the system of equations:
\begin{eqnarray}
\Gamma_{TA}(I_{Smax})\simeq (dI/dt)/I_{c0},\label{ISmax}\\
\Gamma_R(T_{down}) = \Gamma_{TA}.\label{Tdown}
\end{eqnarray}
Here the first equation is the condition for $I_{Smax}$, which
states that the junction should switch into the R-state during the
time of experiment. From Eqs.(\ref{RateTA},\ref{ISmax}) it follows
that $\Delta U (I_{Smax})/k_B T \simeq \ln \left[\frac{a_t
\omega_p I_{c0}}{2 \pi dI/dt} \right] \equiv Y$, which agrees with
experiment, as shown in Fig. 4d). Taking $\Delta U \simeq (4
\sqrt{2}/3) E_{J0}\left[1-I_S/I_{c0}\right]^{3/2}$,
and neglecting $T-$ dependence of $I_{c0}$, we reproduce the
linear $T-$dependence $I_{Smax}/I_{c0}\simeq
1-(3Yk_BT)/(4\sqrt{2}E_{J0})$, seen in Fig.4 a). Substituting this
expression into Eqs.(\ref{Tdown},\ref{RateR}) and assuming
$I_{R0}\simeq\frac{4I_{c0}}{\pi Q_0}$ (valid for $Q_0> 2$), we
obtain:
\begin{equation}
T^* \simeq \frac{16 E_{J0}}{9 Q_0^2 Y^{\frac 1 3}k_B}
\left[\sqrt{1+ \left(1-\frac{4}{\pi Q_0}\right)\frac{3
Q_0^2}{\sqrt{8}Y^{\frac 1 3}}}-1\right]^2. \label{TdownAn}
\end{equation}

From Eq.(\ref{TdownAn}) it follows that $T^*/E_{J0}$ depends
almost solely on $Q_0$. Fig. 4 c) shows $T_{esc}$ vs $T$ for a
S-2DEG-S $\#2^*$ (similar to $\#2$) before and after $C-$shunting.
A dramatic difference in the behavior of TA is obvious. As shown
in Fig. 1a) the $C-$shunting affects almost solely $Q_0$.
Therefore, switching from S to R state is not strongly affected by
$C-$shunting. On the contrary, retrapping is affected considerably
because $I_{R0}\sim 1/Q_0$. Under these circumstances, higher $T$
is required to reduce $I_{Smax}$ to the level of $I_R$, resulting
in the increase of $T^*$.

To get an insight into the phase dynamics at $T>T^*$, we show in
Fig. 4 d) the dependence of $\Delta U(I=I_{Smax})$ vs. $T^*$ for
the case of Fig. 2b). It is seen that $\Delta U$ scales with $T$.
The solid line in Fig. 4d) corresponds to $\Delta U (I_{Smax})
/k_B T = 24.3\simeq Y$ obtained from simulations presented in
Figs. 4 a,b) and demonstrates excellent agreement with experiment.
The large value of $\Delta U /k_BT$ implies that the junction can
escape from S to R state only few times during the time of
experiment. Therefore, the collapse is not due to transition into
the phase-diffusion state, which may also lead to reduction of
$\Delta I$, previously observed in SIS junctions
\cite{Vion,Finish} (even though we assume that the decrease of
$\Delta I$ reported recently in Ref.\cite{Finish} may also be
described by our model). Indeed, phase diffusion requires repeated
escape and retrapping, which is only possible for $\Delta U /k_B T
\simeq 1$ \cite{Kautz,Muller}. Careful measurements of S-branches
in the IVC's at $T \gtrsim T^*$ did not reveal any signature of
dc-voltage down to $\sim 10 nV$ for S-2DEG-S and $\sim 1 \mu V$
for IJJ's. Furthermore, as seen from comparison of Figs. 1b, 2b)
and Figs. 1d, 3c), the IVC's remain hysteretic at $T$ well above
$T^*$, which is incompatible with the phase diffusion according to
the RCSJ model \cite{Kautz}. As can be seen from Fig.1 c) the
first indication for the phase diffusion in our IJJ's appears only
at $T>90K$, meaning that all the collapse of TA shown in Fig. 3 c)
at $75K<T<85K$ occurs before entering into the phase diffusion
state.

For a quantitative comparison with experiment we performed full
numerical simulations of Eqs. (1-3) taking into account the $T-$
dependence of $I_{c0}$, shown in Fig. 1 b) and the exact value of
hysteresis $I_{c0}/I_{R0}$ within the RCSJ model. Results of the
simulations for the S-2DEG-S $\#3$ at four magnetic fields,
corresponding to Figs. 1b) and 2b) are shown in Fig. 2c). Dashed
and solid lines represent the simulated width of histograms
disregarding and taking into account retrapping, respectively. It
is seen that simulations quantitatively reproduce $T^*$ for all
four magnetic fields. The capacitance $C = 0.15 pF$, which was the
only fitting parameter, is the same for all four curves and
corresponds to the expected value of stray capacitance. Taking
into account that $T^*$ is very sensitive to $I_{c0}$ and $C$, see
Eq.(\ref{TdownAn}), we may say that the agreement between theory
and experiment is excellent.

In conclusion, we observed a paradoxical collapse of thermal
activation with increasing $T$ in two very different types of
Josephson junctions with moderate damping. The phenomenon was
explained by the interplay of two conflicting consequences of
thermal fluctuations, which can both assist in premature switching
and help in retrapping back into the S-state. The retrapping
process is significant at small currents, causing cutting-off the
thermal activation at small bias. We have analyzed the influence
of dissipation on the thermal activation by tuning the damping
parameter with the gate voltage, magnetic field, temperature and
in-situ capacitive shunting. Numerical simulations are in good
agreement with experimental data and explain both the paradoxical
collapse and the unusual shape of switching histograms.


\end{document}